\documentclass[preprint,12pt]{elsarticle}

\usepackage{amsmath}
\usepackage{amssymb}
\usepackage{graphicx}
\usepackage{hyperref}

\journal{Proceedings of TOP2025}

\begin{document}

\begin{frontmatter}

\title{State-of-the-art cross sections for $t\bar{t}H$: NNLO+NNLL+EW predictions}

\author{Anna Kulesza}
\address{Institute for Theoretical Physics, University of M\"unster, D-48149 M\"unster, Germany}

\begin{abstract}
The most precise theoretical predictions for the total cross section for the associated production of a Higgs boson with a top-antitop quark pair at the LHC, first presented in \cite{Balsach:2025tth}, are reported. The calculation combines NNLO QCD corrections that include an approximation of the two-loop virtual contribution, with soft-gluon resummation at NNLL accuracy in two independent frameworks (SCET and direct QCD), and complete NLO electroweak corrections. 
\end{abstract}


\end{frontmatter}

\section{Introduction}

The associated production of a Higgs boson with a top-antitop quark pair, $pp \to t\bar{t}H$, provides a direct probe of the top quark Yukawa coupling $y_t$ without making assumptions about its nature, and can be used to study the CP structure of the top-Higgs interaction. This process was first observed in 2018 at the Large Hadron Collider (LHC) by  both ATLAS and CMS collaborations \cite{CMS:2018uxb,ATLAS:2018mme} and has since been extensively studied in various Higgs decay channels. The top-Yukawa coupling, being proportional to mass, represents the strongest Yukawa interaction between fundamental Standard Model (SM) particles, making it particularly relevant for understanding the stability of our Universe as well as searching for physics beyond the SM. 

The reference total cross section in the LHC Higgs Cross Section Working Group Yellow Report 4  was computed at next-to-leading order (NLO) QCD+EW accuracy \cite{deFlorian:2016spz}.
Since then significant theoretical progress has been achieved, including calculations of the next-to-next-to-leading order (NNLO) corrections with suitable approximations of the two-loop amplitudes \cite{Catani:2021cot,Catani:2022mfv,Devoto:2024nhl}, threshold corrections at the next-to-next-to-leading-logarithmic (NNLL) accuracy\cite{Kulesza:2015vda, Broggio:2015lya, Broggio:2016zgg, Broggio:2016lfj,   Broggio:2017kzi, Kulesza:2017ukk, Kulesza:2018tqz,   Broggio:2019ewu, Ju:2019soa, Kulesza:2020nfh}, complete NLO EW corrections \cite{Frixione:2014qaa,Frixione:2015zaa,Frederix:2018nkq}, as well as off-shell effects \cite{Denner:2015yca, Denner:2016kdg, Stremmer:2021bnk}.
This proceedings reports on recent results \cite{Balsach:2025tth} for the $t\bar{t}H$ total cross section  at the LHC that achieve NNLO+NNLL+EW accuracy, providing the most precise theoretical predictions available to date. The calculation builds upon NNLO QCD results \cite{Catani:2021cot,Catani:2022mfv,Devoto:2024nhl}, combines them with NNLL soft-gluon resummation in both direct QCD (dQCD) \cite{Kulesza:2017ukk, Kulesza:2020nfh} and soft-collinear effective theory (SCET) frameworks \cite{Broggio:2016lfj, Broggio:2019ewu}, and includes complete NLO corrections \cite{Frixione:2014qaa,Frixione:2015zaa,Frederix:2018nkq}. A thorough comparison of the NNLO+NNLL results calculated within two different resummation frameworks, dQCD and SCET was carried out, and multiple sources of theoretical uncertainties were carefully analysed. 

\section{NNLO QCD predictions}

The NNLO QCD calculation employs the transverse-momentum ($q_T$) subtraction method \cite{Catani:2007vq}, extending it \cite{Devoto:2025eyc} from the $t\bar{t}$ case \cite{Catani:2019hip}. Although major work towards calculation of exact two-loop amplitudes for the $t \bar t H$ production is ongoing \cite{FebresCordero:2023ixr,Agarwal:2024vvl,Wang:2024lwc}, they still remain unavailable. Correspondingly, results in \cite{Catani:2022mfv,Devoto:2024nhl} rely on applying  approximations to the two-loop amplitudes: the soft Higgs boson approximation and the high-energy limit approximation, calculated using massification procedure. The reader is referred to \cite{Catani:2022mfv,Devoto:2024nhl} for the details of the two approximation methods.

The two-loop contribution is then obtained through a weighted average of results obtained using the two approximations. The NNLO QCD corrections increase the cross section by approximately 4\% relative to NLO QCD. A systematic uncertainty is assigned for each of the approximation schemes, based on the relative discrepancy between exact and approximated predictions at NLO and the effects of subtraction scale $\mu_{\text{IR}}$ variation. The two uncertainties are combined in quadrature, 
yielding $\Delta_{\text{virt}} = 0.9\%$. This error is subdominant compared to perturbative uncertainties at NNLO but must be included in the theory uncertainty budget of matched results.

\section{NNLL threshold resummation}

Soft-gluon resummation addresses large logarithmic corrections arising in the threshold limit $\hat{s} \to Q^2$, where $\hat{s}$ is the partonic center-of-mass energy squared and $Q^2$ is the invariant mass squared of the $t\bar{t}H$ system. Two complementary theoretical frameworks are employed: the dQCD approach \cite{Kulesza:2017ukk, Kulesza:2020nfh} and the SCET approach \cite{Broggio:2016lfj, Broggio:2019ewu}. Although both are based on factorization in the soft limit, the derivation of expressions in the two frameworks differs conceptually. The dQCD expressions are derived from the properties of scattering amplitudes and cross sections in full QCD. In the case of SCET, effective field theory techniques are used in the intermediate steps. These different treatments lead to different organizations of the resummed expressions, so that exact agreement between the two would require including terms to all orders in the power of the logarithm. Correspondingly, when evaluated at the same finite logarithmic accuracy, the analytic and numerical results will differ by contributions beyond the nominal accuracy. For a brief review of the formalisms the reader is referred to \cite{Balsach:2025tth} while more details can be found in \cite{Kulesza:2017ukk, Kulesza:2020nfh, Broggio:2016lfj, Broggio:2019ewu}.

The differences in the implementation of the renormalization group (RG)-improved perturbation theory manifest themselves at the level of analytic expressions in a couple of ways. The dQCD formulas inherit the scale dependence of the fixed-order results, i.e. they depend on the renormalization scale $\mu_R$ and the  factorization scale $\mu_F$. The original construction of the SCET formulas depend on three scales: the hard scale $\mu_h$, the soft scale $\mu_s$ and the  factorization scale $\mu_F$. For the purpose of this work $\mu_s$ was set equal to the value that corresponds to the formulas in the dQCD approach, while the dependence on $\mu_R$ was introduced by eliminating $\alpha_s(\mu_h)$ in favour of $\alpha_s(\mu_R)$, yielding SCET expressions depending on $\mu_R$, $\mu_F$ and $\mu_h$. The scale variation error in SCET is then evaluated by taking an envelope over 11 scale choices, corresponding to the 7-point method used in dQCD extended to accommodate for two choices for $\mu_h$, $\mu_h=\mu_F$ and $\mu_h=\mu_R$. Further, in dQCD all dependence on large logarithms enters through $ \lambda = \alpha_s(\mu_R)\beta_0 \ln N$, where $N$ is the Mellin moment variable of the Mellin transformation taken w.r.t. $Q^2/\hat s$. In contrast, the exponential factors in SCET depend on several order-one parameters of this type, corresponding to logarithms of ratios of any two scales mentioned above. Additionally, specific implementation choices are made. In the SCET approach  the factor $\exp(\alpha_s g_3)$, involving the NNLL function $g_3$, is approximated through its ${\cal O} (\alpha_s)$ expansion, while in the dQCD approach the terms non-vanishing in the $\lambda \to 0$ limit are treated as contributions to the hard function.

The agreement between the resummed expressions in the two approaches up to the nominal logarithmic precision, here NNLL, was verified analytically. Consequently, the implemented expressions differ by terms formally of order N$^3$LL and higher. Thus the numerical differences between the two predictions can be interpreted as an indicator of the size of the subleading effects beyond the formal accuracy of resummation.

The comparison of the numerical results at the NNLO+NNLL accuracy, i.e. including the NNLL resummed correction matched to the NNLO predictions is shown in Fig~\ref{fig1} (left). The NNLO+NNLL predictions agree remarkably well, with the central values differing by only a few permille for all three default scale choices: $\mu_F = \mu_R = (m_t + m_H/2)$, $H_T/2$, and $Q/2$. To account for these differences in the final theoretical predictions the NNLO+NNLL results in dQCD and SCET are combined by averaging over the central values and taking an envelope over the error bands due to scale variation. 

\section{NNLO+NNLL+EW numerical predictions}

The state of the art predictions reported here are obtained by adding complete EW corrections to the matched NNLO+NNLL cross sections. Unless otherwise stated, we use the Higgs mass $m_H = 125.09$ GeV and $\sqrt{S} = 13.6$ TeV, central scale choice $\mu = (m_H + 2m_t)/2$, and other input parameters chosen according to the LHC Higgs Working Group recommendations\footnote{See https://twiki.cern.ch/twiki/bin/view/LHCPhysics/LHCHWG136TeVxsec}.

\begin{figure}
\centering
\includegraphics[width=0.48\linewidth]{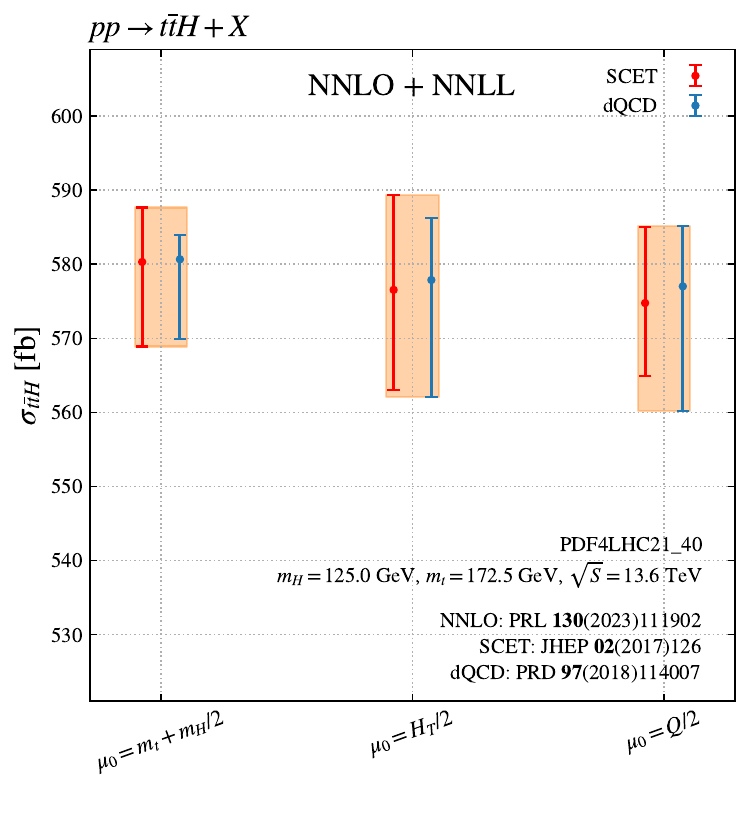}
\includegraphics[width=0.48\textwidth]{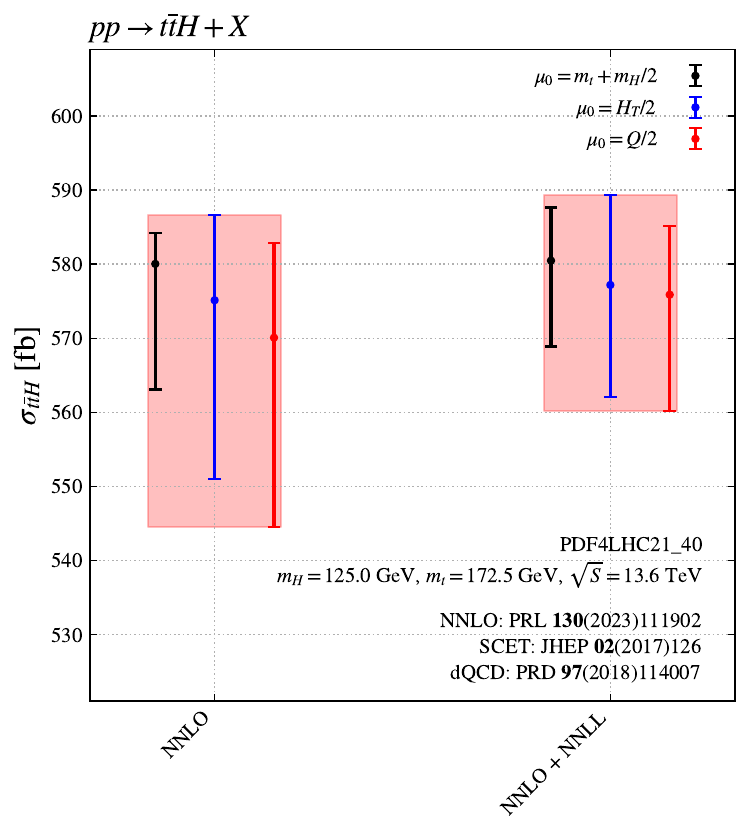}
 \caption{Left: Comparison between NNLO+NNLL results in dQCD and SCET for three parametrically different choices of default scales. Right: Comparison of combined NNLO+NNLL results with NNLO for the same three scale choices. No EW corrections are included.}
\label{fig1}
 \end{figure}

 The impact of supplementing the NNLO predictions with NNLL resummation is shown in Fig.~\ref{fig1}(right). For the default scale choice $\mu = (m_H + 2m_t)/2$ central predictions change only marginally (at the level of 1 permille or below) compared to NNLO. However, for other scale choices, corrections are larger. The EW corrections contribute approximately 2\% to the NNLO total cross section, with negligible impact on scale uncertainties given their small size.

The scale variation uncertainties, estimated as described above, are substantially reduced from approximately 3\% at NNLO to less than 2\% at NNLO+NNLL for different values of the central scale. The NNLO+NNLL predictions are also more stable with respec to the choice of the central default scale than the NNLO results, see Fig.~\ref{fig1}(right).
The scale uncertainties of the full NNLO+NNLL+EW predictions are found to be between 1.5\% and 2.2\% for the default scale choice $\mu = (m_H + 2m_t)/2$. Apart from the scale uncertainties, the dominant sources of theoretical errors are the PDF and $\alpha_s$  uncertainties (determined according to the PDF4LHC prescription) as well the uncertainty stemming from the two loop contribution. Other sources including numerical integration uncertainties, ambiguities in the resummation approach, top-quark mass value, top-quark mass renormalization scheme dependence, and missing higher-order EW corrections are estimated to be subleading at the permille level. Altogether, the state-of-the-art prediction is
\begin{equation}
\sigma_{\text{NNLO+NNLL+EW}} = 592.1~\text{fb}~{}^{+1.5\%}_{-2.2\%}(\Delta_\mu) \pm 2.2\%(\Delta_{\text{PDF}}) \pm 1.7\%(\Delta_{\alpha_s}) \pm 0.9\%(\Delta_{\text{virt}}), \nonumber
\end{equation}
where a parametric uncertainty $\delta\alpha_s = 0.001$ on $\alpha_s$ was assumed. Notably, through the inclusion of NNLO+NNLL corrections and the resulting reduction of the scale error, the theoretical error budget is now dominated by the PDF uncertainties. NNLO+NNLL+EW predictions for other values of $\sqrt S$ and $m_H$ can be found in ~\cite{Balsach:2025tth}.

 \section*{Acknowledgements}

This work was carried out within the LHC Higgs Working Group. The author is grateful to A. Broggio for comments on the manuscript, acknowledges support from the German Research Foundation (DFG) under grants KU 3103/1 and KU 3103/2, and thanks all coauthors of \cite{Balsach:2025tth} for a very pleasant and fruitful collaboration.


\begin{thebibliography}{99}


\bibitem{Balsach:2025tth}
R.~Balsach et al.,
``State-of-the-art cross sections for $t\bar{t}H$: NNLO predictions matched with NNLL resummation and EW corrections,''
arXiv:2503.15043 [hep-ph].

\bibitem{CMS:2018uxb}
CMS Collaboration,
``Observation of $t\bar{t}H$ production,''
Phys. Rev. Lett. \textbf{120}, 231801 (2018).

\bibitem{ATLAS:2018mme}
ATLAS Collaboration,
``Observation of Higgs boson production in association with a top quark pair at the LHC with the ATLAS detector,''
Phys. Lett. B \textbf{784}, 173 (2018).

\bibitem{deFlorian:2016spz}
D.~de Florian et al. [LHC Higgs Cross Section Working Group],
``Handbook of LHC Higgs Cross Sections: 4. Deciphering the Nature of the Higgs Sector,''
arXiv:1610.07922 [hep-ph].



\bibitem{Catani:2021cot}
S.~Catani, I.~Fabre, M.~Grazzini and S.~Kallweit,
``$t\bar{t}H$ production at NNLO: the flavour off-diagonal channels,''
Eur. Phys. J. C \textbf{81}, 491 (2021).

\bibitem{Catani:2022mfv}
S.~Catani, S.~Devoto, M.~Grazzini, S.~Kallweit, J.~Mazzitelli and C.~Savoini,
``Higgs Boson Production in Association with a Top-Antitop Quark Pair in Next-to-Next-to-Leading Order QCD,''
Phys. Rev. Lett. \textbf{130}, 111902 (2023).

\bibitem{Devoto:2024nhl}
S.~Devoto, M.~Grazzini, S.~Kallweit, J.~Mazzitelli and C.~Savoini,
``Precise predictions for $t\bar{t}H$ production at the LHC: inclusive cross section and differential distributions,''
arXiv:2411.15340 [hep-ph].

\bibitem{Kulesza:2015vda}
A.~Kulesza, L.~Motyka, T.~Stebel and V.~Theeuwes,
``Soft gluon resummation for associated $t\bar{t}H$ production at the LHC,''
JHEP \textbf{03}, 065 (2016).




\bibitem{Broggio:2015lya}
A.~Broggio, A.~Ferroglia, B.~D.~Pecjak, A.~Signer and L.~L.~Yang,
``Associated production of a top pair and a Higgs boson beyond NLO,''
JHEP \textbf{03}, 124 (2016).

\bibitem{Broggio:2016zgg}
A.~Broggio, A.~Ferroglia, G.~Ossola and B.~D.~Pecjak,
``Associated production of a top pair and a W boson at next-to-next-to-leading logarithmic accuracy,''
JHEP \textbf{09} (2016), 089.


\bibitem{Broggio:2016lfj}
A.~Broggio, A.~Ferroglia, B.~D.~Pecjak and L.~L.~Yang,
``NNLL resummation for the associated production of a top pair and a Higgs boson at the LHC,''
JHEP \textbf{02}, 126 (2017).

\bibitem{Broggio:2017kzi}
A.~Broggio, A.~Ferroglia, G.~Ossola, B.~D.~Pecjak and R.~D.~Sameshima,
``Associated production of a top pair and a Z boson at the LHC to NNLL accuracy,''
JHEP \textbf{04} (2017), 105.

\bibitem{Kulesza:2017ukk}
A.~Kulesza, L.~Motyka, T.~Stebel and V.~Theeuwes,
``Associated $t\bar{t}H$ production at the LHC: Theoretical predictions at NLO+NNLL accuracy,''
Phys. Rev. D \textbf{97}, 114007 (2018).

\bibitem{Kulesza:2018tqz}
A.~Kulesza, L.~Motyka, D.~Schwartl\"ander, T.~Stebel and V.~Theeuwes,
Associated production of a top quark pair with a heavy electroweak gauge boson at NLO+NNLL accuracy,
Eur. Phys. J. C \textbf{79} (2019) 249.



\bibitem{Broggio:2019ewu}
A.~Broggio, A.~Ferroglia, R.~Frederix, D.~Pagani, B.~D.~Pecjak and I.~Tsinikos,
Top-quark pair hadroproduction in association with a heavy boson at NLO+NNLL including EW corrections,
JHEP \textbf{08} (2019) 039.

\bibitem{Ju:2019soa}
W.-L.~Ju and L.~L.~Yang,
Resummation of soft and Coulomb corrections for tth production at the LHC,
JHEP \textbf{06} (2019) 050.


\bibitem{Kulesza:2020nfh}
A.~Kulesza, L.~Motyka, D.~Schwartl\"ander, T.~Stebel and V.~Theeuwes,
Associated top quark pair production with a heavy boson: differential cross sections at NLO+NNLL accuracy,
Eur. Phys. J. C \textbf{80} (2020) 428.







\bibitem{Frixione:2014qaa}
S.~Frixione, V.~Hirschi, D.~Pagani, H.~S.~Shao and M.~Zaro,
``Weak corrections to Higgs hadroproduction in association with a top-quark pair,''
JHEP \textbf{09}, 065 (2014).

\bibitem{Frixione:2015zaa}
S.~Frixione, V.~Hirschi, D.~Pagani, H.~S.~Shao and M.~Zaro,
``Electroweak and QCD corrections to top-pair hadroproduction in association with heavy bosons,''
JHEP \textbf{06}, 184 (2015).

\bibitem{Frederix:2018nkq}
R.~Frederix, S.~Frixione, V.~Hirschi, D.~Pagani, H.~S.~Shao and M.~Zaro,
``The automation of next-to-leading order electroweak calculations,''
JHEP \textbf{07}, 185 (2018).

\bibitem{Denner:2015yca}
A.~Denner and R.~Feger,
NLO QCD corrections to off-shell top-antitop production with leptonic decays in association with a Higgs boson at the LHC,
JHEP \textbf{11} (2015) 209.

\bibitem{Denner:2016kdg}
A.~Denner, J.-N.~Lang, M.~Pellen and S.~Uccirati,
Higgs production in association with off-shell top-antitop pairs at NLO EW and QCD at the LHC,
JHEP \textbf{02} (2017) 053.

\bibitem{Stremmer:2021bnk}
D.~Stremmer and M.~Worek,
Production and decay of the Higgs boson in association with top quarks,
JHEP \textbf{02} (2022) 196.




\bibitem{Catani:2007vq}
S.~Catani and M.~Grazzini,
``An NNLO subtraction formalism in hadron collisions and its application to Higgs boson production at the LHC,''
Phys. Rev. Lett. \textbf{98}, 222002 (2007).

\bibitem{Devoto:2025eyc}
S.~Devoto and J.~Mazzitelli,
``Soft contributions to heavy quark production in arbitrary kinematics,''
[arXiv:2509.17509 [hep-ph]].

\bibitem{Catani:2019hip}
S.~Catani, S.~Devoto, M.~Grazzini, S.~Kallweit and J.~Mazzitelli,
Top-quark pair hadroproduction at next-to-next-to-leading order in QCD,
Phys. Rev. D \textbf{99} (2019) 051501.

\bibitem{FebresCordero:2023ixr}
F.~Febres Cordero, G.~Figueiredo, M.~Kraus, B.~Page and L.~Reina,
``Two-loop master integrals for leading-color $pp \to t\bar{t}H$ amplitudes with a light-quark loop,''
JHEP \textbf{07}, 084 (2024).

\bibitem{Agarwal:2024vvl}
B.~Agarwal et al.,
``Two-loop amplitudes for $t\bar{t}H$ production: the quark-initiated $N_f$-part,''
JHEP \textbf{05}, 013 (2024).

\bibitem{Wang:2024lwc}
G.~Wang, T.~Xia, L.~L.~Yang and X.~Ye,
``Two-loop QCD amplitudes for $t\bar{t}H$ production from boosted limit,''
JHEP \textbf{07}, 121 (2024).

\end{thebibliography}
\end{document}